\documentclass[article,preprint,amssymb,showpacs,superscriptaddress]{revtex4}

\usepackage{graphicx}
\usepackage{dcolumn}
\usepackage{bm}
\usepackage{color}
\begin{document}


\renewcommand{\baselinestretch}{2} 

\title{Anisotropy of superconducting single crystal SmFeAsO$_{0.8}$F$_{0.2}$ studied by torque magnetometry}

\author{S.~Weyeneth} \email{wstephen@physik.uzh.ch}
\affiliation{Physik-Institut der Universit\"{a}t Z\"{u}rich,
Winterthurerstrasse 190, CH-8057 Z\"urich, Switzerland}

\author{R.~Puzniak}
\affiliation{Institute of Physics, Polish Academy of Sciences, Aleja Lotnik\'ow 32/46, PL 02-668 Warsaw, Poland}

\author{U.~Mosele}
\affiliation{Physik-Institut der Universit\"{a}t Z\"{u}rich,
Winterthurerstrasse 190, CH-8057 Z\"urich, Switzerland}

\author{N.D.~Zhigadlo}
\affiliation{Laboratory for Solid State Physics, ETH Z\"urich, CH-8093
Z\"{u}rich, Switzerland}

\author{S.~Katrych}
\affiliation{Laboratory for Solid State Physics, ETH Z\"urich, CH-8093
Z\"{u}rich, Switzerland}

\author{Z.~Bukowski}
\affiliation{Laboratory for Solid State Physics, ETH Z\"urich, CH-8093
Z\"{u}rich, Switzerland}

\author{J.~Karpinski}
\affiliation{Laboratory for Solid State Physics, ETH Z\"urich, CH-8093
Z\"{u}rich, Switzerland}

\author{S.~Kohout}
\affiliation{Physik-Institut der Universit\"{a}t Z\"{u}rich,
Winterthurerstrasse 190, CH-8057 Z\"urich, Switzerland}

\author{J.~Roos}
\affiliation{Physik-Institut der Universit\"{a}t Z\"{u}rich,
Winterthurerstrasse 190, CH-8057 Z\"urich, Switzerland}

\author{H.~Keller}
\affiliation{Physik-Institut der Universit\"{a}t Z\"{u}rich,
Winterthurerstrasse 190, CH-8057 Z\"urich, Switzerland}

\begin{abstract}
Single crystals of the oxypnictide superconductor SmFeAsO$_{0.8}$F$_{0.2}$ with $T_c\simeq45(1)$ K were investigated by torque magnetometry. The crystals of mass $\leq0.1$ $\mu$g were grown by a high-pressure cubic anvil technique. The use of a high-sensitive piezoresistive torque sensor made it possible to study the anisotropic magnetic properties of these tiny crystals. The anisotropy parameter $\gamma$ was found to be field independent, but varies strongly with temperature ranging from $\gamma\simeq8$ at $T\lesssim T_{c}$ to $\gamma\simeq23$ at $T\simeq 0.4T_{c}$. This unusual behavior of $\gamma$ signals unconventional superconductivity. 
\end{abstract}


\maketitle

\renewcommand{\baselinestretch}{2} 
\section{Introduction}
The search for novel superconductors recently lead to the discovery of the Fe-based oxypnictide layered superconductor LaFeAsO$_{1-x}$F$_x$ with a transition temperature of $T_c\simeq26$ K \cite{Kamihara}. By substituting La with other rare earth ions like Sm, Ce, Nd, Pr, or Gd a series of novel superconducting materials was synthesized with $T_c$'s up to 55~K \cite{Chen1, Chen2, Ren1, Ren2, Ren3, Cheng}. These compounds have a layered crystal structure consisting of LaO and FeAs sheets, where the Fe ions are arranged on simple square lattices \cite{Kamihara}. According to theoretical predictions superconductivity takes place in the FeAs layers, whereas the LaO layers are charge reservoirs when doped with F ions \cite{Singh, Klauss}.\\
The anisotropic behavior of layered superconductors can be characterized by their effective mass anisotropy $\gamma$, which in the framework of the classical anisotropic Ginzburg-Landau theory is given by $\gamma=\sqrt{m^*_{c}/m^*_{ab}}=\lambda_c/\lambda_{ab}=B_{c2}^{ab}/B_{c2}^c$ \cite{Thinkham}. Here $m^*_{ab}$ and $m^*_{c}$ are the effective carrier masses related to supercurrents flowing in the $ab$-planes and along the $c-$axis, respectively, $\lambda_{ab}$ and $\lambda_c$ are the corresponding magnetic penetration depths, and $B_{c2}^{ab}$ and $B_{c2}^{c}$ the corresponding upper critical fields. In order to determine reliable values of $\gamma$ high-quality single crystals are required.\\
An estimation of $\gamma\geq30$ at $T=0$~K was made for \linebreak SmFeAsO$_{0.82}$F$_{0.18}$ from measurements of the $c$-axis plasma frequency using infrared ellipsometry \cite{Bernhard}. From point-contact spectroscopy \cite{Shan} and resistivity experiments \cite{Zhu} on LaFeAsO$_{0.9}$F$_{0.1-\delta}$ the anisotropy parameter was estimated to be $\gamma \simeq 10$. Recent resistivity measurements on single crystal NdFeAsO$_{0.82}$F$_{0.18}$ revealed a value of $\gamma\lesssim5$ \cite{Jia} close to $T_c$. Band structure calculations for LaFeAsO$_{1-x}$F$_{x}$ yield a resistivity anisotropy of $\sim15$ for isotropic scattering, corresponding to $\gamma\simeq4$ \cite{Singh}. For comparison, in MgB$_2$ $\gamma$ ranges from 1 to 8 and shows a strong temperature and field dependence \cite{Angst, Lyard}. This unconventional behavior of $\gamma$ is well described within the framework of two-band superconductivity \cite{Gurevich, Angstgap}.\\
\section{Experiment}
\begin{figure}[b!]
\includegraphics[width=0.75\linewidth]{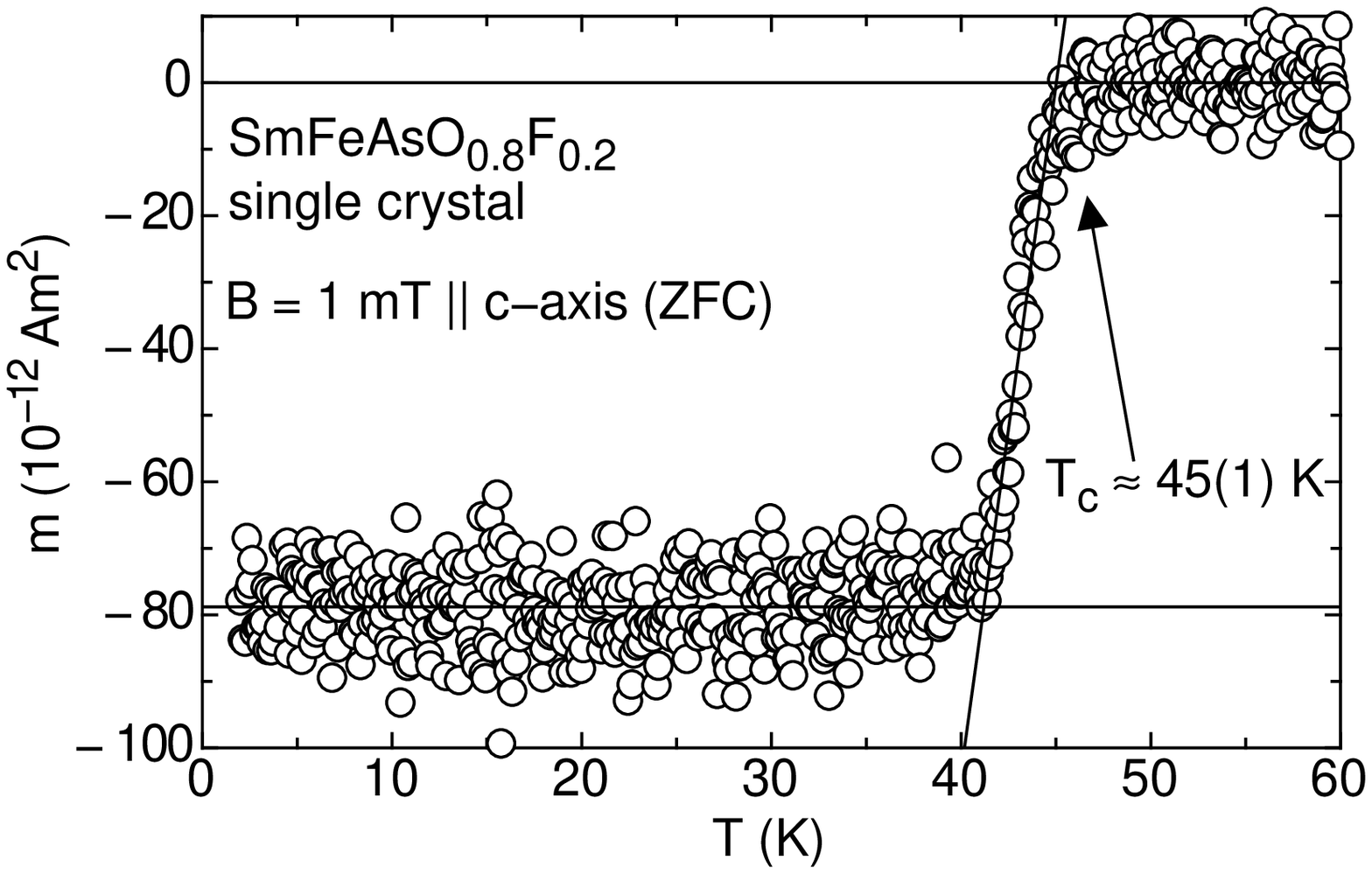}
\vspace{0cm}
\caption{Temperature dependence of the magnetic moment $m$ of a SmFeAsO$_{0.8}$F$_{0.2}$ single crystal measured in 1 mT parallel to the $c$-axis (zero-field cooling, ZFC). The rough estimate of the transition temperature $T_{c}\simeq45(1)$ K (indicated by the arrow) is consistent with the value of $T_{c}=46.0(3)$ K determined from the temperature dependence of the in-plane magnetic penetration depth $\lambda_{ab}$ (see text and Fig.~\ref{fig4}).}
\label{fig1}
\end{figure}

\begin{figure}[t!]
\includegraphics[width=0.75\linewidth]{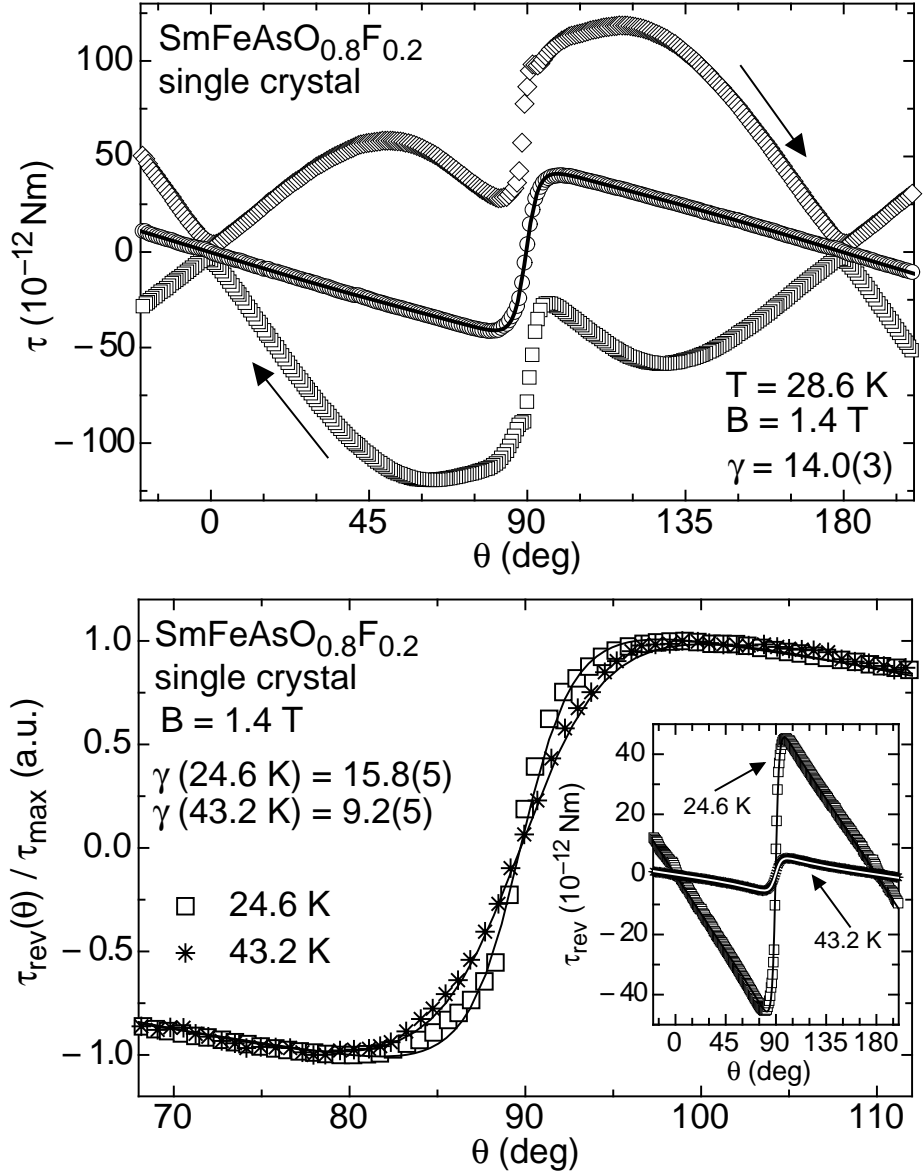}
\vspace{0cm}
\caption{Upper panel: Angular dependence of the reversible torque of a single crystal of SmFeAsO$_{0.8}$F$_{0.2}$ at $T=28.6$ K and $B=1.4$ T. The squares (diamonds) denote a counterclockwise (clockwise) rotating of the magnetic field around the sample. For minimizing pinning effects the mean torque $\tau_{rev}=(\tau(\theta^+)+\tau(\theta^-))/2$ was calculated (circles). The solid line denotes a fit to Eq.~(\ref{kogan}) from which the anisotropy parameter $\gamma=14.0(3)$ was extracted.\\
Lower panel: Normalized angular dependence of the torque for $\theta$ in the neighborhood of 90$^\circ$ ($B$ parallel $ab$-plane) at 24.6 K (squares) and 43.2 K (stars). The solid lines represent fits to Eq.~(\ref{kogan}). The different shapes of the two torque signals reflect the temperature dependence of $\gamma$. The inset displays the full angular dependence of the same torque signals.}
\label{fig2}
\end{figure}

\begin{figure}[t!]
\includegraphics[width=0.75\linewidth]{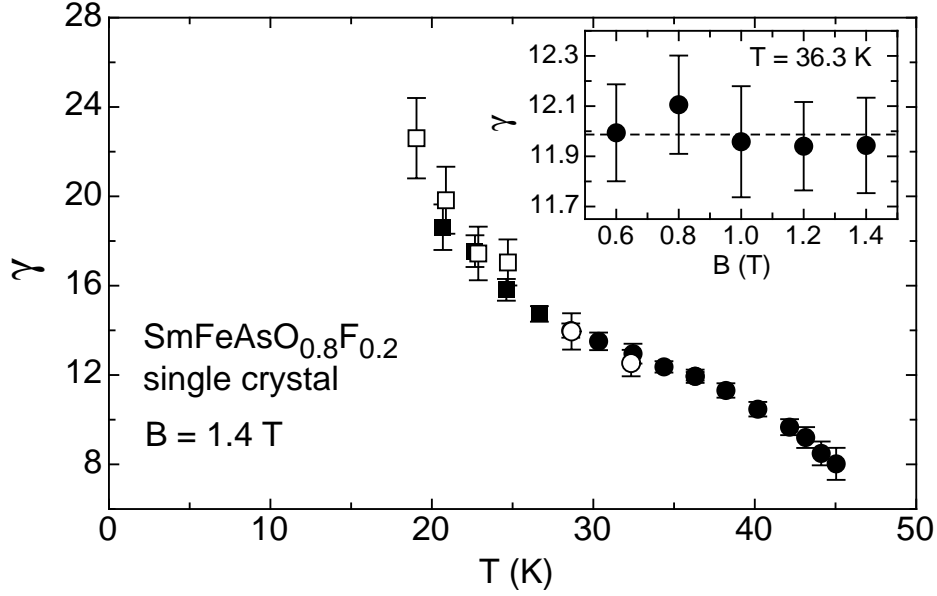}
\vspace{0cm}
\caption{Temperature dependence of the anisotropy parameter $\gamma$ of a single crystal of SmFeAsO$_{0.8}$F$_{0.2}$ at $B=1.4$ T obtained from a subsequent analysis of the torque data by means of Eq.~(\ref{kogan}). Full symbols denote values of $\gamma$ extracted from simply averaged data sets, whereas open symbols show values of $\gamma$ from shaked data sets as explained in the text. The values of $\gamma$ obtained from fits to Eq.~(\ref{kogan}) with all parameters free are indicated by circles, whereas the squares represent the values of $\gamma$ evaluated using extrapolated values of $B_{c2}^c$. The inset shows $\gamma$ determined at 36.3 K from unshaked data at different fields. As can be seen no field dependence of $\gamma$ is observed.}
\label{fig3}
\end{figure}

\begin{figure}[b!]
\includegraphics[width=0.75\linewidth]{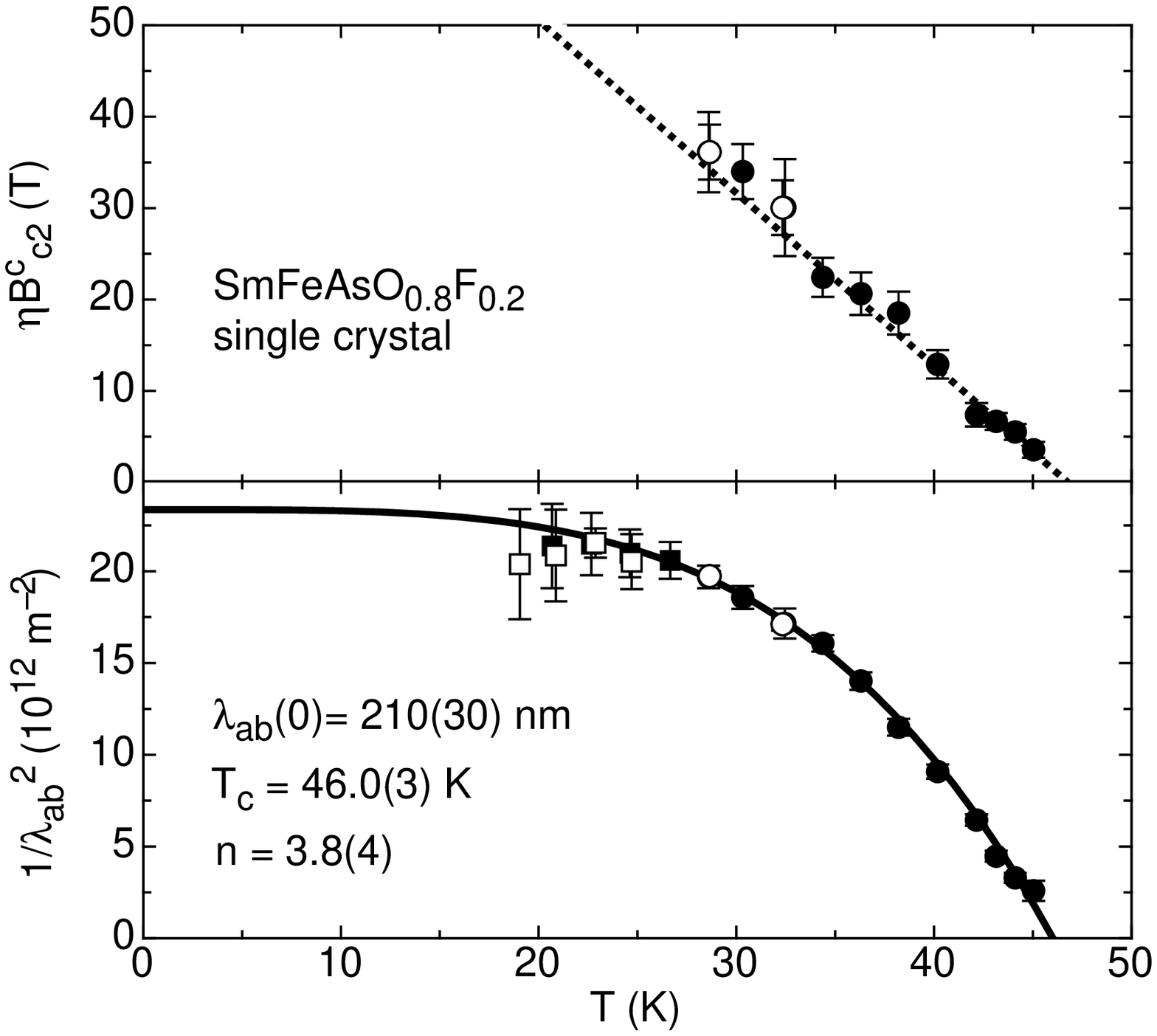}
\vspace{0cm}
\caption{Upper panel: Temperature dependence of the upper critical field $B_{c2}^c$ of a single crystal of SmFeAsO$_{0.8}$F$_{0.2}$. The dashed line is a linear fit $\eta B_{c2}^c=88$ T$ - (1.9$ T/K) $ \cdot T$. The meaning of the symbols is the same as in Fig.~\ref{fig3}.\\
Lower panel: In-plane magnetic penetration depth $\lambda_{ab}(T)$ obtained from the torque data and plotted as $1/\lambda_{ab}^2$ versus temperature. The solid line represents a fit of the data to the power law $1/\lambda_{ab}^2(T)=1/\lambda_{ab}^2(0)\cdot(1-(T/T_c)^n)$ with all parameters free. The exponent of $n=3.8(4)$ is close to $n=4$ characteristic for the two-fluid model. The meaning of the symbols is the same as in Fig.~\ref{fig3}}
\label{fig4}
\end{figure}

Here we report magnetic torque experiments on a single crystal of SmFeAsO$_{0.8}$F$_{0.2}$ with a $T_c\simeq45(1)$ K performed in the temperature range of 19 K to 45 K and in magnetic fields up to 1.4 T. From these measurements the magnetic anisotropy $\gamma$ was extracted in the framework of the anisotropic London model \cite{Kogan1, Kogan2}. Single crystals of nominal composition SmFeAsO$_{0.8}$F$_{0.2}$ \cite{Karpinski2} with masses of only $ \approx 0.1$ $\mu$g were grown from NaCl/KCl flux using a high-pressure cubic anvil technique \cite{Karpinski1, Karpinski3}. The plate-like crystals used in this work are of rectangular shape with typical dimensions of 70$\times$25$\times$3 $\mu$m$^3$, corresponding to an estimated volume of  $V\simeq5\cdot10^{-15}$ m${^3}$. The crystal structure was checked by means of X-ray diffraction revealing the $c$-axis to be perpendicular to the plates.\\
Selected crystals of SmFeAsO$_{0.8}$F$_{0.2}$ of the same batch with  $T_{c} \simeq45$ K were characterized in a commercial {\it Quantum Design Magnetometer} MPMS XL. As an example, Fig.~\ref{fig1} shows the zero-field cooled (ZFC) magnetic moment $m$ of one of the crystals vs. temperature in 1 mT parallel to the $c$-axis. The rough estimate for $T_{c}\simeq45(1)$ K (straight line in Fig.~\ref{fig1}) is consistent with $T_{c}=46.0(3)$ K obtained from the temperature dependence of the in-plane magnetic penetration depth $\lambda_{ab}$ (see Fig.~\ref{fig4}). $m$ saturates in the Meissner state below 40 K and the total transition width is $\Delta T\lesssim4$ K, demonstrating the good quality of the sample.\\
The magnetic torque $\tau$ of a single-crystal sample in a static magnetic field $B$ applied under a certain angle $\theta$ with respect to the crystallographic $c$-axis of the crystal is given by $\vec{\tau}=\vec{m}\times\vec{B}=\mu_0(\vec{m}\times\vec{H})$, or equivalently $\tau(\theta)=m(\theta)B\sin(\phi(\theta))$. Here $\vec{m}$ denotes the magnetic moment of the sample and $\phi(\theta)$ the respective angle between $\vec{B}$ and $\vec{m}$. 
Our apparatus was especially designed for measurements on micro crystals with masses smaller than 1 $\mu$g using micro-machined sensors of high sensitivity ($\sim10^{-14}$ Nm) developed in our group \cite{Kohout}. To collect angular dependent torque data $\tau(\theta)$ we used a measurement system with a turnable {\it Bruker} NMR iron yoke magnet with a maximum magnetic field of 1.4 T, allowing a rotation of the magnetic field direction of more that 360$^\circ$ with respect to a crystallographic axis of the sample. Temperatures down to 10 K can be achieved with a flow cryostat between the poles of the magnet. The possibility of rotating the magnetic field around a fixed sample increases the sensitivity again because background effects to the torque are minimized.\\
Below $T_c$ we observe a torque signal arising from the interaction of the applied magnetic field with vortices in the sample. Based on the 3D anisotropic London model, the free energy of an anisotropic superconductor in the mixed state was calculated by Kogan \textit{et al.} \cite{Kogan1, Kogan2}
\begin{equation}
\tau(\theta)=-\frac{V\Phi_0B}{16\pi\mu_0\lambda_{ab}^2}\bigg(1-\frac{1}{\gamma^2}\bigg)\frac{\sin(2\theta)}{\epsilon(\theta)}\ln\bigg(\frac{\eta B_{c2}^c}{\epsilon(\theta)B}\bigg).
\label{kogan}
\end{equation}
$V$ is the volume of the crystal, $\Phi_0$ is the elementary flux quantum, $B_{c2}^c$ is the upper critical field along the $c$-axis of the crystal, $\eta$ denotes a numerical parameter of the order unity depending on the structure of the flux-line lattice, and $\epsilon(\theta)=[\cos^2(\theta)+\gamma^{-2}\sin^2(\theta)]^{1/2}$. By measuring the angular dependence of the torque in the mixed state of a superconductor, three fundamental parameters can be extracted from the data: the in-plane magnetic penetration depth $\lambda_{ab}$, the $c$-axis upper critical field $B_{c2}^c$, and the effective mass anisotropy $\gamma$ . In this work we performed angle dependent measurements of the torque over more than 180$^\circ$ in order to investigate the full angular dependent magnetization in terms of Eq.~(\ref{kogan}). The torque was measured with a clockwise and a counterclockwise rotating magnetic field and then averaged according to $\tau_{rev}=(\tau(\theta^+)+\tau(\theta^-))/2$ to reduce vortex pinning effects (see upper panel of Fig.~\ref{fig2}). To further minimize the influence of pinning, we partly applied a vortex-lattice shaking technique \cite{Shaking}. With this technique the relaxation of the vortex-lattice towards its thermal equilibrium is strongly enhanced by using an additional small magnetic AC field perpendicular to the static external field $B$. The magnitude of the AC field was $\simeq5$ mT and the shaking frequency was $\simeq$ 200 Hz. With these parameters the best experimental conditions, including the vortex-lattice relaxation towards thermal equilibrium, were obtained. Examples of the reversible torque as a function of the angle $\theta$ determined in the superconducting state are displayed in the lower panel of Fig.~\ref{fig2}. A small temperature independent background of $10^{-12}$ Nm was recorded well above $T_c$ and subtracted from all measurements below $T_c$. \\
The most reliable parameter which can be extracted from the torque data is the anisotropy parameter $\gamma$ because according to Eq.~(\ref{kogan}) it only depends on the angular dependence of the torque and not on the sample volume. Fig.~\ref{fig2} (lower panel) displays two examples of torque measurements performed at $B=1.4$ T and for two different temperatures 24.6 K and 43.2 K, yielding quite different values of $\gamma$, namely $\gamma(24.6$ K$)=15.8(5)$ and $\gamma(43.2$ K$)=9.2(5)$. This is clearly reflected in the different shapes of the torque signals for $\theta$ in the neighborhood of 90$^\circ$ ($B$ almost parallel to the $ab$-plane). The temperature dependence of $\gamma$ determined at 1.4 T is displayed in Fig.~\ref{fig3}. As the temperature decreases $\gamma$ rises from $\gamma(45.0$ K$)=8.0(7)$ to $\gamma(19.0$ K$)=22.6(15)$. Note that the present values of $\gamma$ are in the ballpark of the values so far reported in the literature \cite{Singh, Bernhard, Shan, Zhu, Jia}. At low temperatures pinning effects give rise to the rather large errors in the anisotropy parameter, preventing a reliable determination of $\gamma$ for $T<19$~K. In order to demonstrate that the strong temperature dependence of $\gamma$ is not an artefact of the increasing pinning strength at low temperatures, we compared shaked and unshaked torque data. One would expect that the shaked data are close to the reversible thermal equilibrium state. However, as indicated by the open and closed symbols in Fig.~\ref{fig3} there is within error bars no difference between the values of $\gamma$ determined with and without shaking, attesting that the used procedure to determine $\gamma$ below the irreversibility line is still reliable \cite{Shaking}. Furthermore, the field dependence of $\gamma$ was studied at $T=36.3$ K. As shown in the inset of Fig.~\ref{fig3}, for 0.6 T $\leqslant B\leqslant1.4$ T no field dependence of $\gamma$ could be detected, suggesting that the measured anisotropy parameter in this work is $\gamma=\lambda_{c}/\lambda_{ab}=B_{c2}^{ab}/B_{c2}^{c}$. Since $\gamma$ is field independent its determination is not appreciably affected by weak intrinsic pinning present in the sample.\\
The only parameter which is difficult to extract from Eq.~(\ref{kogan}) in the presence of weak pinning is the upper critical field $B_{c2}^c$ as it enters only logarithmically into the formula. For this reason we fitted $B_{c2}^c(T)$ above 28 K according to $\eta B_{c2}^c(T)\approx88$ T$ - (1.9$ T/K)$\cdot T$ and extrapolated values of $B_{c2}^c(T)$ down to 19 K in order to follow the temperature dependence of $\gamma$ further down in temperature (note that the fitted slope of $dB_{c2}^c/dT\approx-2$ T/K agrees well with the slope close to $T_c$ reported in \cite{Hunte2}). The values obtained from the extrapolated  values of $B_{c2}^c$ are displayed in Fig.~\ref{fig3} as squares, whereas the values evaluated by the free fits are represented by circles.\\
The temperature dependencies of the upper critical field $B_{c2}^c$ and the in-plane magnetic penetration depth $\lambda_{ab}$ as extracted from the torque data are displayed in Fig.~\ref{fig4}. $B_{c2}^c$ follows a linear temperature dependence down to 28 K. The dashed line represents a linear fit in order to extrapolate $B_{c2}^c$ down to 19 K.  Using the WHH approximation \cite{WHH} yields $B_{c2}^c(0)\simeq60$ T. $\lambda_{ab}$ was also estimated from the torque data using Eq.~(\ref{kogan}). In the lower panel of Fig.~\ref{fig4} the temperature dependence of $1/\lambda_{ab}^2$ is displayed. As shown by the solid curve in Fig.~\ref{fig4} (lower panel), the data were analyzed with the power law $1/\lambda_{ab}^2(T)=1/\lambda_{ab}^2(0)\cdot(1-(T/T_c)^n)$ with the free parameters $\lambda_{ab}(0)\simeq210(30)$ nm, $T_c=46.0(3)$ K, and $n=3.8(4)$. This value of $\lambda_{ab}(0)$ is in reasonable agreement with the values reported for polycrystalline samples: From $\mu$SR experiments Luetkens \textit{et al.} extracted $\lambda_{ab}(0)= 254(2)$~nm for LaFeAsO$_{0.9}$F$_{0.1}$ and $\lambda_{ab}(0)= 364(8)$~nm for LaFeAsO$_{0.925}$F$_{0.075}$ \cite{Luetkens}, whereas Khasanov \textit{et al.} obtained  a value of $\lambda_{ab}(0)=198(5)$ nm for SmFeAsO$_{0.85}$ \cite{Khasanov}. The exponent $n=3.8(4)$ is close to $n=4$ for the two-fluid model, characteristic for a superconductor in the strong-coupling limit \cite{Rammer}. However, the present data are too limited to draw definite conclusions.\\
SmFeAsO$_{0.8}$F$_{0.2}$ exhibits a similar temperature dependence of the anisotropy parameter $\gamma$ as MgB$_2$ for which $\gamma$ varies between 1 and 8, depending on temperature and magnetic field  \cite{Angst, Lyard, Angstgap}. Therefore, the unusual temperature dependence of $\gamma$ of SmFeAsO$_{0.8}$F$_{0.2}$, which cannot be explained by conventional Ginzburg-Landau theory, indicates a possible two gap scenario \cite{Angstgap}. Note that other experiments such as recent resistivity measurements of the upper critical field on LaFeAsO$_{0.89}$F$_{0.11}$ \cite{Hunte2} and NMR studies of PrFeAsO$_{0.89}$F$_{0.11}$ \cite{Matano} also suggest that the novel class of Fe-based oxypnictide superconductors exhibit two-band superconductivity. Nevertheless, other possibilities of a  temperature dependent $\gamma$ should be mentioned here \cite{Pugnat, Kawamata, Maki}.\\
\section{Conclusions}
In conclusion, the magnetic anisotropy parameter $\gamma$, and the in-plane magnetic penetration depth $\lambda_{ab}$ were determined by high-sensitive magnetic torque experiments for a single crystal of SmFeAsO$_{0.8}$F$_{0.2}$. The parameter  $\gamma$ extracted from the reversible torque signals was found to be strongly temperature dependent, ranging from $\gamma\simeq8$ at $T\lesssim T_{c}$ to $\gamma\simeq23$ at $T\simeq 0.4T_{c}$. However, no field dependence of $\gamma$ could be detected. For the in-plane magnetic penetration depth at zero temperature a value of $\lambda_{ab}(0) \simeq210(30)$ nm was estimated, in agreement with values reported in the literature \cite{Luetkens, Khasanov}. The unusual temperature dependence of $\gamma$ suggests that SmFeAsO$_{0.8}$F$_{0.2}$ is probably a multi-band superconductor, similar to MgB$_2$ \cite{Angstgap} and the cuprate superconductors \cite{Khas2}. However, other explanations for this unusual behavior such as e.g., the possibility of an anisotropic $s$-wave gap \cite{Maki}, cannot be ruled out. More detailed experimental work is required to clarify this point.
\section{Acknowledgments}
The authors are grateful to A. Bussmann-Holder for very helpful discussions, and to S.~Str\"assle and C.~Duttwyler for the help to prepare the manuscript. This work was supported by the Swiss National Science Foundation and in part by the NCCR program MaNEP and the EU Project 
CoMePhS.

\end{document}